\documentclass[prl,aps,twocolumn,a4paper,showkeys,nofootinbib]{revtex4-1}

\usepackage[colorlinks=true,backref=true,linkcolor=blue,linktocpage=true,anchorcolor=black,citecolor=blue,filecolor=black,menucolor=black,pagecolor=black,urlcolor=blue]{hyperref}

\usepackage{graphicx,psfrag}
\usepackage{mathrsfs}
\usepackage{amsmath,amsfonts,amssymb}
\usepackage{multirow}
\usepackage{comment}
\usepackage{bm}
\usepackage{ulem}
\usepackage{hyperref}

\newcommand{\be}{\begin{equation}}
\newcommand{\ee}{\end{equation}}
\newcommand{\bea}{\begin{eqnarray}}
\newcommand{\eea}{\end{eqnarray}}
\newcommand{\bel}{\begin{align}}
\newcommand{\eel}{\end{align}}

\def\GMc2{G M_{\odot} c^{-2}}

\DeclareSymbolFontAlphabet{\mathrsfs}{rsfs}
\DeclareMathAlphabet{\mathcal}{OMS}{cmsy}{m}{n}
\DeclareSymbolFontAlphabet{\mathrsfs}{rsfs}
\DeclareMathAlphabet\mathbfcal{OMS}{cmsy}{b}{n}

\usepackage{color}
\definecolor{cyan}{rgb}{0,0.9,0.9}
\definecolor{orange}{rgb}{0.9,0.5,0}
\definecolor{magenta}{rgb}{1,0,1}
\definecolor{purple}{rgb}{0.8,0.4,0.8}
\definecolor{gray}{rgb}{0.8242,0.8242,0.8242}

\begin{document}
\title{Empirical tests of the black hole no-hair conjecture using gravitational-wave observations}
\author{Gregorio \surname{Carullo}$^{1,2}$}
\email{gregorio.carullo@ligo.org}
\author{Laura \surname{van der Schaaf}$^{2}$}
\author{Lionel \surname{London}$^{3}$}
\author{Peter T.~H. \surname{Pang}$^{4}$}
\author{Ka~Wa~\surname{Tsang}$^{2}$}
\author{Otto A. \surname{Hannuksela}$^{4}$}
\author{Jeroen \surname{Meidam}$^{2}$}
\author{Michalis \surname{Agathos}$^{5}$}
\author{Anuradha \surname{Samajdar}$^{2}$}
\author{Archisman \surname{Ghosh}$^{2}$}
\author{Tjonnie G.~F. \surname{Li}$^{4}$}
\author{Walter \surname{Del Pozzo}$^{1,6}$}
\author{Chris \surname{Van Den Broeck}$^{2,7}$}

\affiliation{${}^{1}$ Dipartimento di Fisica ``Enrico Fermi'', Universit\`a di Pisa, Pisa I-56127, Italy}
\affiliation{${}^{2}$ Nikhef -- National Institute for Subatomic Physics, Science Park, 1098 XG Amsterdam,
The Netherlands}
\affiliation{${}^{3}$ School of Physics and Astronomy, Cardiff University, The Parade, Cardiff CF24 3AA, UK}
\affiliation{${}^{4}$ Department of Physics, The Chinese University of Hong Kong, Shatin, NT, Hong Kong}
\affiliation{${}^{5}$ DAMTP, Centre for Mathematical Sciences, University of Cambridge, Wilberforce Road, Cambridge CB3 0WA, United Kingdom}
\affiliation{${}^{6}$ INFN sezione di Pisa, Pisa I-56127, Italy}
\affiliation{${}^{7}$ Van Swinderen Institute for Particle Physics and Gravity,
University of Groningen, Nijenborgh 4, 9747 AG Groningen, The Netherlands}%

\date{\today}

\begin{abstract}
We show that second-generation gravitational-wave detectors at their design
sensitivity will allow us to directly probe the ringdown phase of binary black hole
coalescences. This opens the possibility to test the so-called black hole no-hair conjecture in a
statistically rigorous way.
Using state-of-the-art numerical relativity-tuned waveform models and dedicated methods to
effectively isolate the quasi-stationary perturbative regime where a ringdown description is valid, we
demonstrate the capability of measuring the physical parameters of the remnant black hole, and subsequently
determining parameterized deviations from the ringdown of Kerr black holes.
By combining information from $\mathcal{O}(5)$ binary black hole mergers with
realistic signal-to-noise ratios achievable with the current generation of detectors,
the validity of the no-hair conjecture
can be verified with an accuracy of $\sim 1.5\%$ at $90\%$ confidence.
\end{abstract}

\maketitle


\textit{Introduction --} The detection of gravitational waves (GWs) by the LIGO and Virgo Collaborations
\cite{TheVirgo:2014hva, Abbott2018} has opened up a variety of avenues for the observational
exploration of the dynamics of gravity and of the nature of black holes. GW150914~\cite{GW150914}
and subsequent detections
\cite{GW151226,TheLIGOScientific:2016pea,Abbott:2017vtc,Abbott:2017gyy,GW170814,GW170817}
have enabled unique tests of general relativity (GR)
\cite{TGR_paper_GW150914,TheLIGOScientific:2016pea,Abbott:2017vtc,GW170814}.
Among the several detections, GW150914 still holds a special place, not only because it was
the first and the loudest binary black hole event detected, but also because it was the kind of
textbook signal that allowed measurements of the frequency and damping time
of what has been interpreted as the least damped quasi-normal mode (QNM) of the presumed
remnant black hole (BH)
resulting from a binary black hole merger~\cite{TGR_paper_GW150914}.
This sparked considerable interest in the community, since it opened up the prospect of
more in-depth empirical studies of quasi-stationary Kerr black holes \cite{Swetha,Area_test} in the near
future,
as the sensitivity of the Advanced LIGO and Advanced Virgo detectors is progressively improved
\cite{Aasi:2013wya,TheVirgo:2014hva}.
Consistency with the prediction of GR hinted that the end result of GW150914 was indeed a Kerr black
hole \cite{Kerr:1963ud}, but inability to detect more than one QNM did not yet allow tests of some key
GR predictions for these objects.
As first predicted by Vishveshwara~\cite{Vishveshwara} and further investigated by Press~\cite{Press:1971wr},
and Chandrasekhar and Detweiler~\cite{Chandrasekhar:1975zza}, in the regime where linearized general relativity
is valid, the strain of the emitted gravitational-wave signal, at large distances from the BH and
neglecting subdominant power-law tail contributions, takes the form:
\be
h(t) = \sum_{lmn} \mathcal{A}_{lmn}e^{-t/\tau_{lmn}}\cos(\omega_{lmn}t+\phi_{lmn})\,.
\ee
For black holes in GR, all frequencies $\omega_{lmn}$ and
damping times $\tau_{lmn}$ are completely determined by the black hole's mass and
spin.\footnote{BH perturbation
theory alone cannot predict the amplitudes $\mathcal{A}_{lmn}$ and relative phases $\phi_{lmn}$;
in the case of black holes resulting from a binary merger, these are set by the properties of the
parent binary black hole system; see
e.g.~\cite{MMRDNS_paper}.}
This can be viewed as a manifestation of the black hole no-hair conjecture, which essentially
states that in GR, a stationary axisymmetric black hole is determined uniquely by its mass, intrinsic
angular momentum, and electric charge (with the latter expected to be zero for astrophysical
objects) \cite{Ginzburg,Zeldovich,Israel,Carter,Hawking1972,Robinson,Robinson,Mazur,Bunting};
see \cite{Mazur_review} for a review.
This connection is key to several tests that have been proposed in the
literature
\cite{Dreyer:2003bv,Gossan,Ferrari_QNM,Cardoso_QNM_LISA,Meidam, Yang:2017zxs, DaSilvaCosta:2017njq, Thrane:2017lqn, Brito:2018rfr}.
So far the possibility to verify (or refute) experimentally the no-hair conjecture has been explored
mostly in the context of third-generation ground-based~\cite{Punturo:2010zz,Evans:2016mbw} or
space-based~\cite{Danzmann:1996da} gravitational-wave detectors.
In this work, we show that the existing advanced
interferometric detector network, when operating at
design sensitivity, will be capable of testing the no-hair conjecture with
an accuracy of a few percent with the observation of the ringdown signal already for $\mathcal{O}(5)$ GW events.

All the mass quantities quoted in the remainder of the paper are defined in the reference frame of the detector.
These are related to the masses in the source rest frame by a factor of $(1+z)$, with $z$ the source redshift \cite{1987GReGr..19.1163K}.

\textit{Ringdown model --} Our ringdown waveform model is that of Ref.~\cite{MMRDNS_paper},
where a robust method was developed to characterize QNMs up to $l = 5$, including overtones (labelled by the $n$ index),
by making use of numerical relativity (NR) waveforms.
The Weyl scalar $\psi_4$ can be expanded as:
\begin{equation}
\psi_4(\iota, \varphi, r, t)
\simeq \frac{M}{r} \sum_{l, m, n} \psi_{lmn}(t)\,\left[{}_{-2}S_{lm}(a_f\tilde{\omega}_{lmn},\iota,\varphi)\right],
\label{psi4}
\end{equation}
with
\begin{equation}
\psi_{lmn}(t) \equiv A_{lmn} e^{i\tilde{\omega}_{lmn} t}.
\end{equation}
In the above, $r$ is the distance from source to detector,  $(\iota, \varphi)$ give the orientation
of the ringing black hole with respect to the line of sight and ${}_{-2}S_{lm}$ are spin-weighted spheroidal harmonics.
For the dependence of the complex mode frequencies $\tilde\omega \equiv \omega_{lmn} + i/\tau_{lmn}$ on
the mass $M_f$ and dimensionless spin $a_f$ of the final black
hole one can use the expressions from \cite{Cardoso_QNM_LISA}.
The \textit{amplitudes} $A_{lmn}$ of the various modes are set by the properties of the initial
black holes that gave rise to the remnant object. As shown in \cite{MMRDNS_paper}, in
the case of non-spinning progenitor objects, these are well captured by series expansions in the
symmetric mass ratio $\eta \equiv m_1 m_2/(m_1+m_2)^2$, with $m_1$, $m_2$ the individual masses.
The coefficients in these expansions are obtained by fitting against NR waveforms
starting from a time $t = 10\,M$ after the peak luminosity of the $(2,2)$ component
of $\psi_4^{\rm NR}$, with $M = m_1 + m_2$; for details of the fitting procedure we refer
to \cite{MMRDNS_paper}. In this setup, the $A_{lmn}$ are complex, so that they
include relative phases between the modes, which were neglected in previous models \cite{Kamaretsos} and
subsequent Bayesian analyses that were based on them \cite{Gossan,Meidam}; their inclusion
leads to a significant improvement in faithfulness against NR waveforms \cite{MMRDSNP}.
At large distances from the source, the gravitational-wave polarizations $h_+$, $h_\times$ are obtained from Eq.~(\ref{psi4})
through $\psi_4 \simeq \ddot{h}_+ -i \ddot{h}_\times$. The waveform model has recently
been extended to the case of
initial black holes with non-zero but aligned spins \cite{MMRDSNP}. In this work we want to provide
a proof of principle that linearized general relativity around a Kerr background can be
directly probed with gravitational-wave observations with the current interferometric network, and the non-spinning model of \cite{MMRDNS_paper} suffices to demonstrate this.\\

\textit{Simulations --} Both to establish the effective ringdown start time and in
subsequent simulations of no-hair conjecture tests, Bayesian parameter estimation is performed.
The simulated signals are numerical inspiral-merger-ringdown
waveforms taken from the publicly available Simulating eXtreme Spacetimes (SXS) catalog \cite{SXS_catalog, Blackman:2015pia}, with mass ratio
$q = m_1/m_2$ in the interval $\left[ 1,3 \right]$, and negligible initial spins as well as negligible residual
eccentricity (SXS:BBH:0001, SXS:BBH:0030, SXS:BBH:0169, SXS:BBH:0198).
These are coherently injected into synthetic, stationary, Gaussian noise for a network of Advanced LIGO and Advanced Virgo
detectors at design sensitivity \cite{Aasi:2013wya,TheVirgo:2014hva}. The injected
total mass is uniformly distributed in the interval $[50, 90]\,M_\odot$, and the sky position as well as
the orientation of the orbital plane at some reference time are uniformly distributed on the sphere.
Luminosity distances $D_L$ are chosen such that the total signal-to-noise ratio (SNR) in the inspiral-merger-ringdown
signal approximately equals 100, which is a plausible value for signals similar to GW150914 \cite{GW150914}
assuming the Advanced LIGO-Virgo network at full sensitivity.
This sets the average SNR contained just in the ringdown phase of our dataset to
15, if the start time is chosen to be $16\, M$ after the time at which the GW strain peaks (as will be demonstrated below, this is indeed a reasonable choice). 
By comparison, with the same choice of start time, the SNR in the ringdown of GW150914 with
detectors at design sensitivity would have been $\text{SNR}_{\rm ring} \simeq 17$ \cite{TGR_paper_GW150914}.

The template waveforms used in our Bayesian analyses follow the aforementioned ringdown model,
augmented with a windowing procedure for the start time, as explained below. Sampling is done over
10 parameters:
\begin{equation}\label{params}
\{ M_f, a_f, q, \alpha, \delta, \iota, \psi, D_L, t_c, \varphi_c \},
\end{equation}
where $(\alpha, \delta)$ determine the sky location, $\psi$ is the polarization angle and $t_c$, $\varphi_c$ respectively are a reference
time and phase. Hence only parameters associated with the ringdown waveform are sampled over, with
the mass ratio $q$ determining the mode amplitudes. Bayesian inference is done using the
\texttt{LALInference} library \cite{LALInference_paper}.
Priors are chosen to be uniform in $[5, 200]\,M_\odot$ for $M_f$,
uniform in $[-1,1]$ for $a_f$, and uniform in $[1,15]$ for $q$.
A constant number density in comoving volume sets the prior for
the sky location angles and the distance, with a distance range of $[1, 1000]$ Mpc.
The priors on $(\iota,\psi)$ are chosen to be uniform on the 2-sphere (with these angles 
being generated from the same distribution also for the simulation set).
(For the SNRs considered, the impact of the specific shapes of the prior distributions has little impact 
on our results.)  
The time of coalescence is uniformly distributed within $[t_c-0.1\,\mbox{s},t_c+0.1\,\mbox{s}]$,
where $t_c$ is a reference time at which the signal is detected.

\textit{When does the ringdown start? --} The time at which the transition between
the non-linear to the linear regime happens is not well defined.
For instance, Ref.~\cite{TGR_paper_GW150914} shows how the inference on the
QNM central frequency and characteristic time changes quite dramatically depending on
the assumed time at which the transition occurs. Therefore it is critical to make a reasonable choice for
the time at which the remnant black hole can be treated perturbatively and assess the effectiveness
of such a choice.

\begin{figure}[t]
\center
\includegraphics[width=0.5\textwidth]{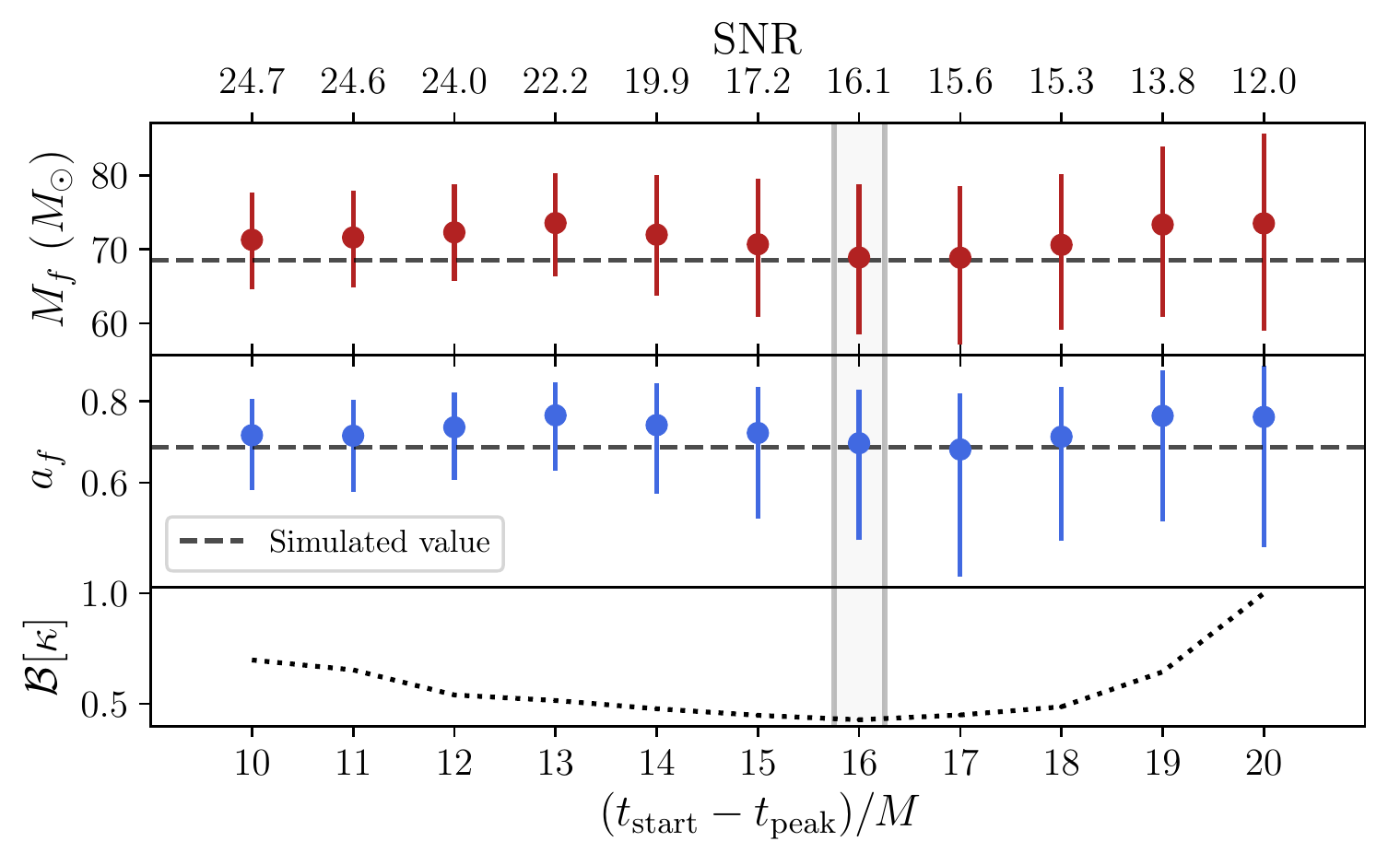}
\caption{\label{fig:q_2_M_65} Estimated median values and 90\% credible regions for final mass $M_f$ (top panel) and
spin $a_f$ (central panel) as a function of the start time of the ringdown with respect to the
strain peak time $t_{\rm peak}$ for a simulation with $M_f=68.5\,M_\odot$ and $a_f=0.686$ (corresponding to $q=1$).
The bottom panel reports the value of the function $\mathcal{B}(k)$, averaged over \textit{all} simulations. The gray box highlights the value of
$t_{\rm start} - t_{\rm peak} = 16\,M$
for which $\mathcal{B}(k)$ is at a minimum.}
\end{figure}

We choose the start time for the ringdown $t_{\rm start}$ from the analysis of numerical
inspiral-merger-ringdown waveforms added to stationary Gaussian noise with a power spectral density
as expected for Advanced LIGO and Virgo at design sensitivity~\cite{noise_curves}. To
isolate the ringdown, we apply to the data a Planck~\cite{Planck_window} window 
whose starting time is varied in discrete steps over a range $[10, 20]\,M$ after the peak time of the strain in each
detector, with $M$ the total mass for the simulated signal\footnote{The choice of a specific 
windowing function has no significant impact on the analysis as long as the frequency range of 
interest is not altered. 
We indeed verified that different tapering functions give nearly identical results.}.
The peak time $t_\text{peak}$ itself can be estimated using analysis methods that can measure amplitudes and arrival times
of a signal inside a detector, without relying on specific GR waveform models, such
as \texttt{BayesWave}~\cite{BayesWave}.
We choose a rise time for the Planck window  of $1$ ms, as we find that this choice
gives a good compromise between the need to preserve the signal-to-noise-ratio and to avoid Gibbs phenomena.
Consider a simulated signal with total mass $M$ and mass ratio $q$, and a choice of window start time,
e.g.~$t_{\rm start} = t_\text{peak} + \kappa M$ for some $\kappa \in [10,20]$. We then apply
a similar window on the ringdown template waveforms, letting them start at $\kappa\, M'$
after the peak strain, where the mass $M'$ is obtained from the \textit{sampled values} $M'_f$ and $q'$ through fitting formulae ~\cite{UIB2016_FF_paper}.
This leads to posterior density functions for all parameters, and in particular for
$M_f$ and $a_f$.
Our criterion to select the start time for the ringdown is built by 
minimizing the bias in the recovered parameters of the final object, 
while avoiding to select an arbitrarily large start time.
Although large start times would ensure the validity of the linearized 
approximation employed in the waveform template, they would also drastically
reduce the signal SNR, thus resulting in a poor estimation of the final parameters.
The equilibrium point in this trade-off, arrived at as explained below, will ensure the analysis
to take place in the linearized regime where our model is valid, 
while still allowing a precise estimation of the measured parameters.
By looking at the covariance between $M_f$ and $a_f$ and at the 
distance (induced by the covariance metric) between the true values $M^I_f$, $a^I_f$ and the mean values $\bar{M}^I_f(\kappa)$ and $\bar{a}^I_f(\kappa)$,
for each simulated signal $I$, we define the functions:
\begin{align}\label{eq:statistics}
\mathcal{B}(\kappa, I) &\equiv \sqrt{D^2_I(\kappa)+ \det\mathbf{C(\kappa)}_I},\\
D^2_I(\kappa) &\equiv \,^t\Delta \vec{x}(\kappa)_I~\mathbf{C}^{-1}\mathbf{(\kappa)}_I~\Delta \vec{x}(\kappa)_I
\end{align}
where $\mathbf{C(\kappa)}_I$ is the two-dimensional covariance matrix between the posterior samples for $M_f$ and $a_f$, $D_I(\kappa)$ is the (covariance induced) distance between the mean and the injected values and we defined the vector $^t\Delta\vec{x}(\kappa)_I = \left(\,(\bar{M}^I_f(\kappa)-M^I_f)/M_{\odot}, \,\bar{a}^I_f(\kappa)-a^I_f \, \right)$.
The statistical uncertainty (larger for large start times), quantified by $\det\mathbf{C(\kappa)}_I$, is controlled by the signal-to-noise left in the ringdown part of the signal when the preceding stages are cut from the analysis. 
The distance $D_I(\kappa)$ quantifies the systematic uncertainty in the 
recovered parameters and is a proxy for the mismatch between the linear 
ringdown model and the non-linear signal.
We thus let the effective ringdown start time be the one that minimizes $\mathcal{B}(\kappa)$ 
(thus minimizing the combination of statistical and systematic uncertainties),
defined as the average of $\mathcal{B}(\kappa, I)$ over all simulated signals. 
The dataset consisted of 12 simulations at 11 different, equally spaced,
values of $\kappa \in [10,20]$, thus employing a total of 132 simulations. 
Figure~\ref{fig:q_2_M_65} illustrates the procedure by
showing $90\%$ credible regions for $M_f$ and $a_f$, together with the value of the \textit{averaged} $\mathcal{B}(\kappa)$ as a function of the ringdown start time, for a particular system which had initial total mass $M = 72\,M_\odot$ and mass ratio $q = 1$.
The value of $\kappa$ minimizing $\mathcal{B}(\kappa)$ is $\kappa = 16$, which implies an
effective ringdown start time of $t_{\rm start} = t_\text{peak} + 16\, M$
after the peak strain of the signal. This is consistent with the conclusions stated in a 
independent study by Bhagwat et al., using a ``Kerrness" measure on a single GW150914-like numerical 
signal~\cite{Swetha}.
The selection of the ringdown start time at $t_{\rm start} = t_\text{peak} + 16\, M$ uniquely 
determines the placement of the time-domain Planck window.
When dealing with real signals, the window is initially applied once to the data with 
$t_{\rm start} = t_\text{peak} + 16\, M_\text{IMR}$, with $t_\text{peak}$ from a model-independent reconstruction as explained above, and $M_\text{IMR}$
from a routine estimate (before performing our ringdown-only analysis) using an inspiral-merger-ringdown model\footnote{By studying a set of numerical simulations, we verified that different definitions of 
the fixed start time of the data window have negligible impact on our results, since this choice just sets the amount of SNR being excluded from the analysis.}.
While the posterior distribution for $M$ (obtained from the \textit{sampled} values of $M_f$ and $q$ through fitting formulae) is being explored,
the window on the template model is instead recalculated 
and reapplied at each step, with its starting time set to the proposed value $16\, M$ after the peak strain.

\begin{figure}[t]
\center
\includegraphics[width=0.5\textwidth]{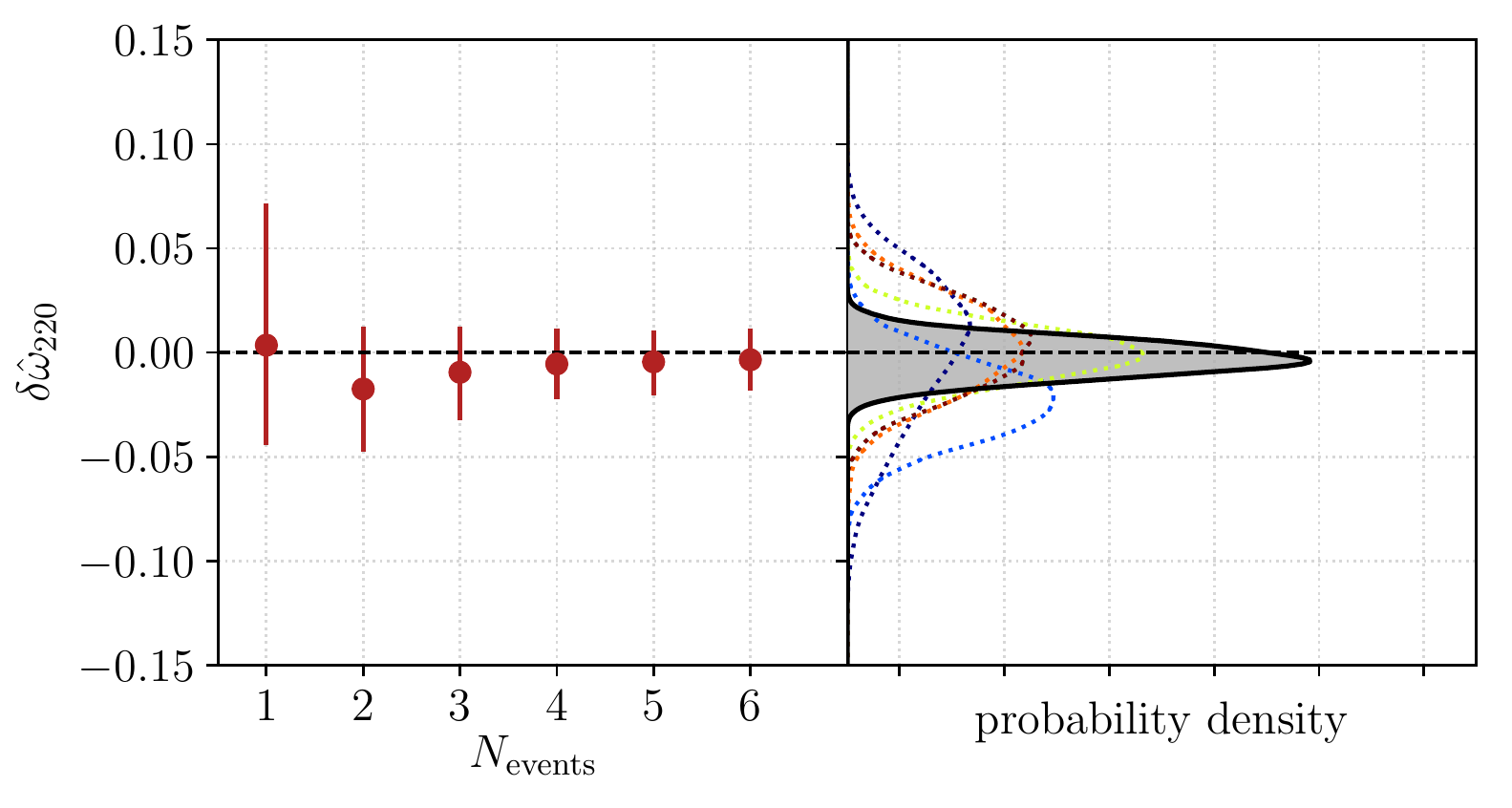}
\caption{\label{fig:domega220} Measurement of the departure of $\delta\hat\omega_{220}$ characterizing a departure of the dominant QNM frequency from its GR value, on a set of numerical simulations as described in the text. \textit{Left panel:}  evolution of medians and 90\% credible intervals from the joint posterior distribution. \textit{Right panel:} posterior probability densities for each individual signal (dotted lines), and the joint posterior distribution (solid line).
With $6$ detections the upper bound on $\delta\hat\omega_{220}$ is smaller than $\sim 1.5\%$ at $90\%$ confidence.}
\end{figure}

\textit{Testing the no-hair conjecture --} As introduced in Refs.~\cite{Gossan,Meidam} 
in the context of third-generation detectors, we look for violations of the
black hole no-hair conjecture by introducing linear deviations in the QNM parameters. In particular, we perturb around the
QNM frequencies and damping times as predicted by GR as:
\begin{eqnarray}\label{eq:violation}
\omega_{lmn}(M_f, a_f) &\rightarrow&  (1+\delta\hat\omega_{lmn})\,\omega_{lmn}(M_f, a_f) \nonumber\\
\tau_{lmn}(M_f, a_f) &\rightarrow& (1+\delta\hat\tau_{lmn})\,\tau_{lmn}(M_f, a_f)
\end{eqnarray}
where $\delta\hat\omega_{lmn}$ and $\delta\hat\tau_{lmn}$ are relative deviations that we include
as additional degrees of freedom in our inference. The parameterization in (\ref{eq:violation}) has the advantages
of being agnostic to specific families of violations and, most importantly, to be uniquely defined in GR,
$\delta\hat\omega_{lmn} = \delta\hat\tau_{lmn} \equiv 0\,\, \forall\, \, l,m,n$.
The no-hair conjecture constrains the number of independent degrees of
freedom of an axisymmetric black hole in GR to be two; therefore,
the conjecture is tested by measuring at least three independent parameters characterizing the remnant geometry,
which we chose to be $M_f, a_f, \delta\hat{\omega}_{220}$.
In addition, the algorithm also samples all the other parameters as specified above.
The priors are unchanged except on $M_f, a_f$ where we restrict to values
contained in the $90\%$ credible intervals obtained from a earlier analysis including inspiral and merger.
The prior on $\delta\hat{\omega}_{220}$ is chosen to be uniform in [-1,1].
We consider GR signals with mass ratio $q \leq 3$, for which the
best QNM determined parameter is the frequency for the least-damped $(2,2,0)$ mode. We thus
focus on the accuracy of the measurement of $\delta\hat\omega_{220}$.
Figure~\ref{fig:domega220} shows the results of an analysis performed
on a set of inspiral-merger-ringdown signals added to stationary Gaussian noise as described above.
Upper bounds on the departures from the predictions of GR for $\omega_{220}$
are smaller than $\sim 1.5\%$ at the $90\%$ credible level
already with six sources, while upper bounds on deviations from $\tau_{220}$ 
predictions are smaller than $\mathcal{O}(10\%)$.
On the selected dataset higher modes deviations on both frequency and damping time
are essentially unconstrained.


\textit{Conclusions --} In this work, we demonstrated that observationally
testing the black hole no-hair conjecture is possible within the next few years, once
the LIGO-Virgo detector network reaches its design sensitivity.
The ability to isolate the quasi-linear ringdown regime from the non-linear merger stage of the coalescence process
enables estimating the parameters characterizing the ringdown.
This also allows the identification of the time of the transition to be $16\,M$ after the peak strain, $M$ being the total mass of the merging
system. Following our procedure, we showed that, with just $\mathcal{O}(5)$ plausible signals,
violations from the no-hair conjecture, seen as changes in the dominant QNM frequency 
and damping time, can be constrained to be smaller than, respectively, $\sim 1.5\%$ and ~$\sim 10\%$ at $90\%$ confidence.
The results presented in this work can be extended to the recent spin-aligned ringdown model 
from Ref.~\cite{MMRDSNP}. This and results for actual signals in LIGO/Virgo data are deferred to a later 
publication.

\textit{Addendum --}
While this work was being finalized, a study by Brito and collaborators~\cite{Brito:2018rfr}
appeared as a pre-print. Although the method proposed
is substantially different, their conclusions are similar to ours.

\begin{acknowledgments}
\textit{Acknowledgments}
G.C. would like to thank Alessandro Nagar for helpful discussions.
The authors would like to thank Maria Haney for a critical reading of the manuscript and for valuable suggestions towards its improvement.
L.v.d.S, K.W.T., J.M., A.S., A.G., and C.V.D.B.~are supported by the
research program of the Netherlands Organisation for Scientific Research (NWO).
P.T.H.P, O.A.H., T.G.F.L. are partially supported by a grant from the Research Grants Council 
of the Hong Kong (Project No. CUHK 24304317) and the Direct Grant for Research 
from the Research Committee of the Chinese University of Hong Kong
O.A.H. is also supported by the the Hong Kong PhD Fellowship Scheme (HKPFS) 
issued by the Research Grants Council (RGC) of Hong Kong.
M.A.~acknowledges NWO-Rubicon Grant No.~RG86688. W.D.P. is funded by the
``Rientro dei Cervelli Rita Levi Montalcini'' Grant of the Italian MIUR.
We are grateful for computational resources provided by Cardiff University, and
funded by an STFC grant supporting UK Involvement in the Operation of Advanced
LIGO. This work greatly benefited from discussions within the \textit{strong-field} working group
of the LIGO/Virgo collaboration.\\
\end{acknowledgments}

\bibliographystyle{apsrev}
\bibliography{No_hair_test_Bibliography}

\begin{thebibliography}{49}
\expandafter\ifx\csname natexlab\endcsname\relax\def\natexlab#1{#1}\fi
\expandafter\ifx\csname bibnamefont\endcsname\relax
  \def\bibnamefont#1{#1}\fi
\expandafter\ifx\csname bibfnamefont\endcsname\relax
  \def\bibfnamefont#1{#1}\fi
\expandafter\ifx\csname citenamefont\endcsname\relax
  \def\citenamefont#1{#1}\fi
\expandafter\ifx\csname url\endcsname\relax
  \def\url#1{\texttt{#1}}\fi
\expandafter\ifx\csname urlprefix\endcsname\relax\def\urlprefix{URL }\fi
\providecommand{\bibinfo}[2]{#2}
\providecommand{\eprint}[2][]{\url{#2}}

\bibitem[{\citenamefont{Acernese et~al.}(2015)}]{TheVirgo:2014hva}
\bibinfo{author}{\bibfnamefont{F.}~\bibnamefont{Acernese}} \bibnamefont{et~al.}
  (\bibinfo{collaboration}{Virgo}), \bibinfo{journal}{Class. Quant. Grav.}
  \textbf{\bibinfo{volume}{32}}, \bibinfo{pages}{024001}
  (\bibinfo{year}{2015}), \eprint{1408.3978}.

\bibitem[{\citenamefont{Abbott et~al.}(2018)\citenamefont{Abbott, Abbott,
  Abbott, Abernathy, Acernese, Ackley, Adams, Adams, Addesso, Adhikari
  et~al.}}]{Abbott2018}
\bibinfo{author}{\bibfnamefont{B.~P.} \bibnamefont{Abbott}},
  \bibinfo{author}{\bibfnamefont{R.}~\bibnamefont{Abbott}},
  \bibinfo{author}{\bibfnamefont{T.~D.} \bibnamefont{Abbott}},
  \bibinfo{author}{\bibfnamefont{M.~R.} \bibnamefont{Abernathy}},
  \bibinfo{author}{\bibfnamefont{F.}~\bibnamefont{Acernese}},
  \bibinfo{author}{\bibfnamefont{K.}~\bibnamefont{Ackley}},
  \bibinfo{author}{\bibfnamefont{C.}~\bibnamefont{Adams}},
  \bibinfo{author}{\bibfnamefont{T.}~\bibnamefont{Adams}},
  \bibinfo{author}{\bibfnamefont{P.}~\bibnamefont{Addesso}},
  \bibinfo{author}{\bibfnamefont{R.~X.} \bibnamefont{Adhikari}},
  \bibnamefont{et~al.}, \bibinfo{journal}{Living Reviews in Relativity}
  \textbf{\bibinfo{volume}{21}}, \bibinfo{pages}{3} (\bibinfo{year}{2018}),
  ISSN \bibinfo{issn}{1433-8351},
  \urlprefix\url{https://doi.org/10.1007/s41114-018-0012-9}.

\bibitem[{\citenamefont{et~al.}(2016{\natexlab{a}})}]{GW150914}
\bibinfo{author}{\bibfnamefont{A.}~\bibnamefont{et~al.}}
  (\bibinfo{collaboration}{LIGO Scientific Collaboration and Virgo
  Collaboration}), \bibinfo{journal}{Phys. Rev. Lett.}
  \textbf{\bibinfo{volume}{116}}, \bibinfo{pages}{061102}
  (\bibinfo{year}{2016}{\natexlab{a}}),
  \urlprefix\url{https://link.aps.org/doi/10.1103/PhysRevLett.116.061102}.

\bibitem[{\citenamefont{et~al.}(2016{\natexlab{b}})}]{GW151226}
\bibinfo{author}{\bibfnamefont{A.}~\bibnamefont{et~al.}}
  (\bibinfo{collaboration}{LIGO Scientific Collaboration and Virgo
  Collaboration}), \bibinfo{journal}{Phys. Rev. Lett.}
  \textbf{\bibinfo{volume}{116}}, \bibinfo{pages}{241103}
  (\bibinfo{year}{2016}{\natexlab{b}}),
  \urlprefix\url{https://link.aps.org/doi/10.1103/PhysRevLett.116.241103}.

\bibitem[{\citenamefont{Abbott et~al.}(2016)}]{TheLIGOScientific:2016pea}
\bibinfo{author}{\bibfnamefont{B.~P.} \bibnamefont{Abbott}}
  \bibnamefont{et~al.} (\bibinfo{collaboration}{Virgo, LIGO Scientific}),
  \bibinfo{journal}{Phys. Rev.} \textbf{\bibinfo{volume}{X6}},
  \bibinfo{pages}{041015} (\bibinfo{year}{2016}), \eprint{1606.04856}.

\bibitem[{\citenamefont{Abbott et~al.}(2017{\natexlab{a}})}]{Abbott:2017vtc}
\bibinfo{author}{\bibfnamefont{B.~P.} \bibnamefont{Abbott}}
  \bibnamefont{et~al.} (\bibinfo{collaboration}{VIRGO, LIGO Scientific}),
  \bibinfo{journal}{Phys. Rev. Lett.} \textbf{\bibinfo{volume}{118}},
  \bibinfo{pages}{221101} (\bibinfo{year}{2017}{\natexlab{a}}),
  \eprint{1706.01812}.

\bibitem[{\citenamefont{Abbott et~al.}(2017{\natexlab{b}})}]{Abbott:2017gyy}
\bibinfo{author}{\bibfnamefont{B.~P.} \bibnamefont{Abbott}}
  \bibnamefont{et~al.} (\bibinfo{collaboration}{Virgo, LIGO Scientific}),
  \bibinfo{journal}{Astrophys. J.} \textbf{\bibinfo{volume}{851}},
  \bibinfo{pages}{L35} (\bibinfo{year}{2017}{\natexlab{b}}),
  \eprint{1711.05578}.

\bibitem[{\citenamefont{et~al.}(2017{\natexlab{a}})}]{GW170814}
\bibinfo{author}{\bibfnamefont{A.}~\bibnamefont{et~al.}}
  (\bibinfo{collaboration}{LIGO Scientific Collaboration and Virgo
  Collaboration}), \bibinfo{journal}{Phys. Rev. Lett.}
  \textbf{\bibinfo{volume}{119}}, \bibinfo{pages}{141101}
  (\bibinfo{year}{2017}{\natexlab{a}}),
  \urlprefix\url{https://link.aps.org/doi/10.1103/PhysRevLett.119.141101}.

\bibitem[{\citenamefont{et~al.}(2017{\natexlab{b}})}]{GW170817}
\bibinfo{author}{\bibfnamefont{A.}~\bibnamefont{et~al.}}
  (\bibinfo{collaboration}{LIGO Scientific Collaboration and Virgo
  Collaboration}), \bibinfo{journal}{Phys. Rev. Lett.}
  \textbf{\bibinfo{volume}{119}}, \bibinfo{pages}{161101}
  (\bibinfo{year}{2017}{\natexlab{b}}),
  \urlprefix\url{https://link.aps.org/doi/10.1103/PhysRevLett.119.161101}.

\bibitem[{\citenamefont{et~al.}(2016{\natexlab{c}})}]{TGR_paper_GW150914}
\bibinfo{author}{\bibfnamefont{A.}~\bibnamefont{et~al.}}
  (\bibinfo{collaboration}{LIGO Scientific and Virgo Collaborations}),
  \bibinfo{journal}{Phys. Rev. Lett.} \textbf{\bibinfo{volume}{116}},
  \bibinfo{pages}{221101} (\bibinfo{year}{2016}{\natexlab{c}}),
  \urlprefix\url{https://link.aps.org/doi/10.1103/PhysRevLett.116.221101}.

\bibitem[{\citenamefont{Bhagwat et~al.}(2017)\citenamefont{Bhagwat, Okounkova,
  Ballmer, Brown, Giesler, Scheel, and Teukolsky}}]{Swetha}
\bibinfo{author}{\bibfnamefont{S.}~\bibnamefont{Bhagwat}},
  \bibinfo{author}{\bibfnamefont{M.}~\bibnamefont{Okounkova}},
  \bibinfo{author}{\bibfnamefont{S.~W.} \bibnamefont{Ballmer}},
  \bibinfo{author}{\bibfnamefont{D.~A.} \bibnamefont{Brown}},
  \bibinfo{author}{\bibfnamefont{M.}~\bibnamefont{Giesler}},
  \bibinfo{author}{\bibfnamefont{M.~A.} \bibnamefont{Scheel}},
  \bibnamefont{and} \bibinfo{author}{\bibfnamefont{S.~A.}
  \bibnamefont{Teukolsky}} (\bibinfo{year}{2017}), \eprint{1711.00926}.

\bibitem[{\citenamefont{Cabero et~al.}(2017)\citenamefont{Cabero, Capano,
  Fischer-Birnholtz, Krishnan, Nielsen, and Nitz}}]{Area_test}
\bibinfo{author}{\bibfnamefont{M.}~\bibnamefont{Cabero}},
  \bibinfo{author}{\bibfnamefont{C.~D.} \bibnamefont{Capano}},
  \bibinfo{author}{\bibfnamefont{O.}~\bibnamefont{Fischer-Birnholtz}},
  \bibinfo{author}{\bibfnamefont{B.}~\bibnamefont{Krishnan}},
  \bibinfo{author}{\bibfnamefont{A.~B.} \bibnamefont{Nielsen}},
  \bibnamefont{and} \bibinfo{author}{\bibfnamefont{A.~H.} \bibnamefont{Nitz}}
  (\bibinfo{year}{2017}), \eprint{1711.09073}.

\bibitem[{\citenamefont{Abbott et~al.}(2013)}]{Aasi:2013wya}
\bibinfo{author}{\bibfnamefont{B.~P.} \bibnamefont{Abbott}}
  \bibnamefont{et~al.} (\bibinfo{collaboration}{VIRGO, LIGO Scientific})
  (\bibinfo{year}{2013}), \bibinfo{note}{[Living Rev. Rel.19,1(2016)]},
  \eprint{1304.0670}.

\bibitem[{\citenamefont{Kerr}(1963)}]{Kerr:1963ud}
\bibinfo{author}{\bibfnamefont{R.~P.} \bibnamefont{Kerr}},
  \bibinfo{journal}{Phys. Rev. Lett.} \textbf{\bibinfo{volume}{11}},
  \bibinfo{pages}{237} (\bibinfo{year}{1963}).

\bibitem[{\citenamefont{Vishveshwara}(1970)}]{Vishveshwara}
\bibinfo{author}{\bibfnamefont{C.~V.} \bibnamefont{Vishveshwara}},
  \bibinfo{journal}{Phys. Rev. D} \textbf{\bibinfo{volume}{1}},
  \bibinfo{pages}{2870} (\bibinfo{year}{1970}),
  \urlprefix\url{https://link.aps.org/doi/10.1103/PhysRevD.1.2870}.

\bibitem[{\citenamefont{Press}(1971)}]{Press:1971wr}
\bibinfo{author}{\bibfnamefont{W.~H.} \bibnamefont{Press}},
  \bibinfo{journal}{Astrophys. J.} \textbf{\bibinfo{volume}{170}},
  \bibinfo{pages}{L105} (\bibinfo{year}{1971}).

\bibitem[{\citenamefont{Chandrasekhar and
  Detweiler}(1975)}]{Chandrasekhar:1975zza}
\bibinfo{author}{\bibfnamefont{S.}~\bibnamefont{Chandrasekhar}}
  \bibnamefont{and} \bibinfo{author}{\bibfnamefont{S.~L.}
  \bibnamefont{Detweiler}}, \bibinfo{journal}{Proc. Roy. Soc. Lond.}
  \textbf{\bibinfo{volume}{A344}}, \bibinfo{pages}{441} (\bibinfo{year}{1975}).

\bibitem[{\citenamefont{London et~al.}(2014)\citenamefont{London, Shoemaker,
  and Healy}}]{MMRDNS_paper}
\bibinfo{author}{\bibfnamefont{L.}~\bibnamefont{London}},
  \bibinfo{author}{\bibfnamefont{D.}~\bibnamefont{Shoemaker}},
  \bibnamefont{and} \bibinfo{author}{\bibfnamefont{J.}~\bibnamefont{Healy}},
  \bibinfo{journal}{Phys. Rev. D} \textbf{\bibinfo{volume}{90}},
  \bibinfo{pages}{124032} (\bibinfo{year}{2014}),
  \urlprefix\url{https://link.aps.org/doi/10.1103/PhysRevD.90.124032}.

\bibitem[{\citenamefont{Ginzburg and Ozernoy}(1964)}]{Ginzburg}
\bibinfo{author}{\bibfnamefont{V.}~\bibnamefont{Ginzburg}} \bibnamefont{and}
  \bibinfo{author}{\bibfnamefont{L.}~\bibnamefont{Ozernoy}},
  \bibinfo{journal}{Zh. Eksp. Teor. Fiz.} \textbf{\bibinfo{volume}{147}},
  \bibinfo{pages}{1030} (\bibinfo{year}{1964}).

\bibitem[{\citenamefont{Doroshkevic et~al.}(1965)\citenamefont{Doroshkevic,
  Zeldovich, and Novikov}}]{Zeldovich}
\bibinfo{author}{\bibfnamefont{A.}~\bibnamefont{Doroshkevic}},
  \bibinfo{author}{\bibfnamefont{Y.~B.} \bibnamefont{Zeldovich}},
  \bibnamefont{and} \bibinfo{author}{\bibfnamefont{I.}~\bibnamefont{Novikov}},
  \bibinfo{journal}{Sov. Phys. JETP 36 1 (7)}  (\bibinfo{year}{1965}).

\bibitem[{\citenamefont{Israel}(1967)}]{Israel}
\bibinfo{author}{\bibfnamefont{W.}~\bibnamefont{Israel}},
  \bibinfo{journal}{Phys. Rev.} \textbf{\bibinfo{volume}{164}},
  \bibinfo{pages}{1776} (\bibinfo{year}{1967}),
  \urlprefix\url{https://link.aps.org/doi/10.1103/PhysRev.164.1776}.

\bibitem[{\citenamefont{Carter}(1971)}]{Carter}
\bibinfo{author}{\bibfnamefont{B.}~\bibnamefont{Carter}},
  \bibinfo{journal}{Phys. Rev. Lett.} \textbf{\bibinfo{volume}{26}},
  \bibinfo{pages}{331} (\bibinfo{year}{1971}),
  \urlprefix\url{https://link.aps.org/doi/10.1103/PhysRevLett.26.331}.

\bibitem[{\citenamefont{Hawking}(1972)}]{Hawking1972}
\bibinfo{author}{\bibfnamefont{S.~W.} \bibnamefont{Hawking}},
  \bibinfo{journal}{Communications in Mathematical Physics}
  \textbf{\bibinfo{volume}{25}}, \bibinfo{pages}{152} (\bibinfo{year}{1972}),
  ISSN \bibinfo{issn}{1432-0916},
  \urlprefix\url{https://doi.org/10.1007/BF01877517}.

\bibitem[{\citenamefont{Robinson}(1975)}]{Robinson}
\bibinfo{author}{\bibfnamefont{D.~C.} \bibnamefont{Robinson}},
  \bibinfo{journal}{Phys. Rev. Lett.} \textbf{\bibinfo{volume}{34}},
  \bibinfo{pages}{905} (\bibinfo{year}{1975}),
  \urlprefix\url{https://link.aps.org/doi/10.1103/PhysRevLett.34.905}.

\bibitem[{\citenamefont{Mazur}(1982)}]{Mazur}
\bibinfo{author}{\bibfnamefont{P.~O.} \bibnamefont{Mazur}},
  \bibinfo{journal}{J. Phys.} \textbf{\bibinfo{volume}{A15}},
  \bibinfo{pages}{3173} (\bibinfo{year}{1982}).

\bibitem[{\citenamefont{Bunting}(1983)}]{Bunting}
\bibinfo{author}{\bibfnamefont{G.}~\bibnamefont{Bunting}},
  \bibinfo{journal}{Ph. D. Thesis (unpublished) University of New England,
  Armidale, N. S. W.}  (\bibinfo{year}{1983}).

\bibitem[{\citenamefont{Mazur}(2000)}]{Mazur_review}
\bibinfo{author}{\bibfnamefont{P.~O.} \bibnamefont{Mazur}}
  (\bibinfo{year}{2000}), \eprint{hep-th/0101012}.

\bibitem[{\citenamefont{Dreyer et~al.}(2004)\citenamefont{Dreyer, Kelly,
  Krishnan, Finn, Garrison, and Lopez-Aleman}}]{Dreyer:2003bv}
\bibinfo{author}{\bibfnamefont{O.}~\bibnamefont{Dreyer}},
  \bibinfo{author}{\bibfnamefont{B.~J.} \bibnamefont{Kelly}},
  \bibinfo{author}{\bibfnamefont{B.}~\bibnamefont{Krishnan}},
  \bibinfo{author}{\bibfnamefont{L.~S.} \bibnamefont{Finn}},
  \bibinfo{author}{\bibfnamefont{D.}~\bibnamefont{Garrison}}, \bibnamefont{and}
  \bibinfo{author}{\bibfnamefont{R.}~\bibnamefont{Lopez-Aleman}},
  \bibinfo{journal}{Class. Quant. Grav.} \textbf{\bibinfo{volume}{21}},
  \bibinfo{pages}{787} (\bibinfo{year}{2004}), \eprint{gr-qc/0309007}.

\bibitem[{\citenamefont{Gossan et~al.}(2012)\citenamefont{Gossan, Veitch, and
  Sathyaprakash}}]{Gossan}
\bibinfo{author}{\bibfnamefont{S.}~\bibnamefont{Gossan}},
  \bibinfo{author}{\bibfnamefont{J.}~\bibnamefont{Veitch}}, \bibnamefont{and}
  \bibinfo{author}{\bibfnamefont{B.~S.} \bibnamefont{Sathyaprakash}},
  \bibinfo{journal}{Phys. Rev. D} \textbf{\bibinfo{volume}{85}},
  \bibinfo{pages}{124056} (\bibinfo{year}{2012}),
  \urlprefix\url{https://link.aps.org/doi/10.1103/PhysRevD.85.124056}.

\bibitem[{\citenamefont{Ferrari and Gualtieri}(2008)}]{Ferrari_QNM}
\bibinfo{author}{\bibfnamefont{V.}~\bibnamefont{Ferrari}} \bibnamefont{and}
  \bibinfo{author}{\bibfnamefont{L.}~\bibnamefont{Gualtieri}},
  \bibinfo{journal}{Gen. Rel. Grav.} \textbf{\bibinfo{volume}{40}},
  \bibinfo{pages}{945} (\bibinfo{year}{2008}), \eprint{0709.0657}.

\bibitem[{\citenamefont{Berti et~al.}(2006)\citenamefont{Berti, Cardoso, and
  Will}}]{Cardoso_QNM_LISA}
\bibinfo{author}{\bibfnamefont{E.}~\bibnamefont{Berti}},
  \bibinfo{author}{\bibfnamefont{V.}~\bibnamefont{Cardoso}}, \bibnamefont{and}
  \bibinfo{author}{\bibfnamefont{C.~M.} \bibnamefont{Will}},
  \bibinfo{journal}{Phys. Rev.} \textbf{\bibinfo{volume}{D73}},
  \bibinfo{pages}{064030} (\bibinfo{year}{2006}), \eprint{gr-qc/0512160}.

\bibitem[{\citenamefont{Meidam et~al.}(2014)\citenamefont{Meidam, Agathos, Van
  Den~Broeck, Veitch, and Sathyaprakash}}]{Meidam}
\bibinfo{author}{\bibfnamefont{J.}~\bibnamefont{Meidam}},
  \bibinfo{author}{\bibfnamefont{M.}~\bibnamefont{Agathos}},
  \bibinfo{author}{\bibfnamefont{C.}~\bibnamefont{Van Den~Broeck}},
  \bibinfo{author}{\bibfnamefont{J.}~\bibnamefont{Veitch}}, \bibnamefont{and}
  \bibinfo{author}{\bibfnamefont{B.~S.} \bibnamefont{Sathyaprakash}},
  \bibinfo{journal}{Phys. Rev. D} \textbf{\bibinfo{volume}{90}},
  \bibinfo{pages}{064009} (\bibinfo{year}{2014}),
  \urlprefix\url{https://link.aps.org/doi/10.1103/PhysRevD.90.064009}.

\bibitem[{\citenamefont{Yang et~al.}(2017)\citenamefont{Yang, Yagi, Blackman,
  Lehner, Paschalidis, Pretorius, and Yunes}}]{Yang:2017zxs}
\bibinfo{author}{\bibfnamefont{H.}~\bibnamefont{Yang}},
  \bibinfo{author}{\bibfnamefont{K.}~\bibnamefont{Yagi}},
  \bibinfo{author}{\bibfnamefont{J.}~\bibnamefont{Blackman}},
  \bibinfo{author}{\bibfnamefont{L.}~\bibnamefont{Lehner}},
  \bibinfo{author}{\bibfnamefont{V.}~\bibnamefont{Paschalidis}},
  \bibinfo{author}{\bibfnamefont{F.}~\bibnamefont{Pretorius}},
  \bibnamefont{and} \bibinfo{author}{\bibfnamefont{N.}~\bibnamefont{Yunes}},
  \bibinfo{journal}{Phys. Rev. Lett.} \textbf{\bibinfo{volume}{118}},
  \bibinfo{pages}{161101} (\bibinfo{year}{2017}), \eprint{1701.05808}.

\bibitem[{\citenamefont{Da~Silva~Costa
  et~al.}(2017)\citenamefont{Da~Silva~Costa, Tiwari, Klimenko, and
  Salemi}}]{DaSilvaCosta:2017njq}
\bibinfo{author}{\bibfnamefont{C.~F.} \bibnamefont{Da~Silva~Costa}},
  \bibinfo{author}{\bibfnamefont{S.}~\bibnamefont{Tiwari}},
  \bibinfo{author}{\bibfnamefont{S.}~\bibnamefont{Klimenko}}, \bibnamefont{and}
  \bibinfo{author}{\bibfnamefont{F.}~\bibnamefont{Salemi}}
  (\bibinfo{year}{2017}), \eprint{1711.00551}.

\bibitem[{\citenamefont{Thrane et~al.}(2017)\citenamefont{Thrane, Lasky, and
  Levin}}]{Thrane:2017lqn}
\bibinfo{author}{\bibfnamefont{E.}~\bibnamefont{Thrane}},
  \bibinfo{author}{\bibfnamefont{P.~D.} \bibnamefont{Lasky}}, \bibnamefont{and}
  \bibinfo{author}{\bibfnamefont{Y.}~\bibnamefont{Levin}},
  \bibinfo{journal}{Phys. Rev.} \textbf{\bibinfo{volume}{D96}},
  \bibinfo{pages}{102004} (\bibinfo{year}{2017}), \eprint{1706.05152}.

\bibitem[{\citenamefont{Brito et~al.}(2018)\citenamefont{Brito, Buonanno, and
  Raymond}}]{Brito:2018rfr}
\bibinfo{author}{\bibfnamefont{R.}~\bibnamefont{Brito}},
  \bibinfo{author}{\bibfnamefont{A.}~\bibnamefont{Buonanno}}, \bibnamefont{and}
  \bibinfo{author}{\bibfnamefont{V.}~\bibnamefont{Raymond}}
  (\bibinfo{year}{2018}), \eprint{1805.00293}.

\bibitem[{\citenamefont{Punturo et~al.}(2010)}]{Punturo:2010zz}
\bibinfo{author}{\bibfnamefont{M.}~\bibnamefont{Punturo}} \bibnamefont{et~al.},
  \bibinfo{journal}{Class. Quant. Grav.} \textbf{\bibinfo{volume}{27}},
  \bibinfo{pages}{194002} (\bibinfo{year}{2010}).

\bibitem[{\citenamefont{Abbott et~al.}(2017{\natexlab{c}})}]{Evans:2016mbw}
\bibinfo{author}{\bibfnamefont{B.~P.} \bibnamefont{Abbott}}
  \bibnamefont{et~al.} (\bibinfo{collaboration}{LIGO Scientific}),
  \bibinfo{journal}{Class. Quant. Grav.} \textbf{\bibinfo{volume}{34}},
  \bibinfo{pages}{044001} (\bibinfo{year}{2017}{\natexlab{c}}),
  \eprint{1607.08697}.

\bibitem[{\citenamefont{Danzmann}(1996)}]{Danzmann:1996da}
\bibinfo{author}{\bibfnamefont{K.}~\bibnamefont{Danzmann}},
  \bibinfo{journal}{Class. Quant. Grav.} \textbf{\bibinfo{volume}{13}},
  \bibinfo{pages}{A247} (\bibinfo{year}{1996}).

\bibitem[{\citenamefont{{Krolak} and {Schutz}}(1987)}]{1987GReGr..19.1163K}
\bibinfo{author}{\bibfnamefont{A.}~\bibnamefont{{Krolak}}} \bibnamefont{and}
  \bibinfo{author}{\bibfnamefont{B.~F.} \bibnamefont{{Schutz}}},
  \bibinfo{journal}{General Relativity and Gravitation}
  \textbf{\bibinfo{volume}{19}}, \bibinfo{pages}{1163} (\bibinfo{year}{1987}).

\bibitem[{\citenamefont{Kamaretsos et~al.}(2012)\citenamefont{Kamaretsos,
  Hannam, Husa, and Sathyaprakash}}]{Kamaretsos}
\bibinfo{author}{\bibfnamefont{I.}~\bibnamefont{Kamaretsos}},
  \bibinfo{author}{\bibfnamefont{M.}~\bibnamefont{Hannam}},
  \bibinfo{author}{\bibfnamefont{S.}~\bibnamefont{Husa}}, \bibnamefont{and}
  \bibinfo{author}{\bibfnamefont{B.~S.} \bibnamefont{Sathyaprakash}},
  \bibinfo{journal}{Phys. Rev. D} \textbf{\bibinfo{volume}{85}},
  \bibinfo{pages}{024018} (\bibinfo{year}{2012}),
  \urlprefix\url{https://link.aps.org/doi/10.1103/PhysRevD.85.024018}.

\bibitem[{\citenamefont{London}(2018)}]{MMRDSNP}
\bibinfo{author}{\bibfnamefont{L.~T.} \bibnamefont{London}}
  (\bibinfo{year}{2018}), \eprint{1801.08208}.

\bibitem[{\citenamefont{Mroue et~al.}(2013)\citenamefont{Mroue, Scheel,
  Szilagyi, Pfeiffer, Boyle, Hemberger, Kidder, Lovelace, Ossokine, Taylor
  et~al.}}]{SXS_catalog}
\bibinfo{author}{\bibfnamefont{A.}~\bibnamefont{Mroue}},
  \bibinfo{author}{\bibfnamefont{M.}~\bibnamefont{Scheel}},
  \bibinfo{author}{\bibfnamefont{B.}~\bibnamefont{Szilagyi}},
  \bibinfo{author}{\bibfnamefont{H.}~\bibnamefont{Pfeiffer}},
  \bibinfo{author}{\bibfnamefont{M.}~\bibnamefont{Boyle}},
  \bibinfo{author}{\bibfnamefont{D.}~\bibnamefont{Hemberger}},
  \bibinfo{author}{\bibfnamefont{L.}~\bibnamefont{Kidder}},
  \bibinfo{author}{\bibfnamefont{G.}~\bibnamefont{Lovelace}},
  \bibinfo{author}{\bibfnamefont{S.}~\bibnamefont{Ossokine}},
  \bibinfo{author}{\bibfnamefont{N.}~\bibnamefont{Taylor}},
  \bibnamefont{et~al.}, \bibinfo{journal}{Phys. Rev. Lett.}
  \textbf{\bibinfo{volume}{111}}, \bibinfo{pages}{241104}
  (\bibinfo{year}{2013}), \eprint{1304.6077}.

\bibitem[{\citenamefont{Blackman et~al.}(2015)\citenamefont{Blackman, Field,
  Galley, Szilágyi, Scheel, Tiglio, and Hemberger}}]{Blackman:2015pia}
\bibinfo{author}{\bibfnamefont{J.}~\bibnamefont{Blackman}},
  \bibinfo{author}{\bibfnamefont{S.~E.} \bibnamefont{Field}},
  \bibinfo{author}{\bibfnamefont{C.~R.} \bibnamefont{Galley}},
  \bibinfo{author}{\bibfnamefont{B.}~\bibnamefont{Szilágyi}},
  \bibinfo{author}{\bibfnamefont{M.~A.} \bibnamefont{Scheel}},
  \bibinfo{author}{\bibfnamefont{M.}~\bibnamefont{Tiglio}}, \bibnamefont{and}
  \bibinfo{author}{\bibfnamefont{D.~A.} \bibnamefont{Hemberger}},
  \bibinfo{journal}{Phys. Rev. Lett.} \textbf{\bibinfo{volume}{115}},
  \bibinfo{pages}{121102} (\bibinfo{year}{2015}), \eprint{1502.07758}.

\bibitem[{\citenamefont{Veitch et~al.}(2015)\citenamefont{Veitch, Raymond,
  Farr, Farr, Graff, Vitale, Aylott, Blackburn, Christensen, Coughlin
  et~al.}}]{LALInference_paper}
\bibinfo{author}{\bibfnamefont{J.}~\bibnamefont{Veitch}},
  \bibinfo{author}{\bibfnamefont{V.}~\bibnamefont{Raymond}},
  \bibinfo{author}{\bibfnamefont{B.}~\bibnamefont{Farr}},
  \bibinfo{author}{\bibfnamefont{W.}~\bibnamefont{Farr}},
  \bibinfo{author}{\bibfnamefont{P.}~\bibnamefont{Graff}},
  \bibinfo{author}{\bibfnamefont{S.}~\bibnamefont{Vitale}},
  \bibinfo{author}{\bibfnamefont{B.}~\bibnamefont{Aylott}},
  \bibinfo{author}{\bibfnamefont{K.}~\bibnamefont{Blackburn}},
  \bibinfo{author}{\bibfnamefont{N.}~\bibnamefont{Christensen}},
  \bibinfo{author}{\bibfnamefont{M.}~\bibnamefont{Coughlin}},
  \bibnamefont{et~al.}, \bibinfo{journal}{Phys. Rev. D}
  \textbf{\bibinfo{volume}{91}}, \bibinfo{pages}{042003}
  (\bibinfo{year}{2015}),
  \urlprefix\url{https://link.aps.org/doi/10.1103/PhysRevD.91.042003}.

\bibitem[{noi()}]{noise_curves}
\emph{\bibinfo{title}{Advanced {LIGO} anticipated sensitivity curves}},
  \bibinfo{note}{{L}IGO Document T0900288-v3},
  \urlprefix\url{https://dcc.ligo.org/LIGO-T0900288/public}.

\bibitem[{\citenamefont{McKechan et~al.}(2010)\citenamefont{McKechan, Robinson,
  and Sathyaprakash}}]{Planck_window}
\bibinfo{author}{\bibfnamefont{D.~J.~A.} \bibnamefont{McKechan}},
  \bibinfo{author}{\bibfnamefont{C.}~\bibnamefont{Robinson}}, \bibnamefont{and}
  \bibinfo{author}{\bibfnamefont{B.~S.} \bibnamefont{Sathyaprakash}},
  \bibinfo{journal}{Class. Quant. Grav.} \textbf{\bibinfo{volume}{27}},
  \bibinfo{pages}{084020} (\bibinfo{year}{2010}), \eprint{1003.2939}.

\bibitem[{\citenamefont{Cornish and Littenberg}(2015)}]{BayesWave}
\bibinfo{author}{\bibfnamefont{N.~J.} \bibnamefont{Cornish}} \bibnamefont{and}
  \bibinfo{author}{\bibfnamefont{T.~B.} \bibnamefont{Littenberg}},
  \bibinfo{journal}{Class. Quant. Grav.} \textbf{\bibinfo{volume}{32}},
  \bibinfo{pages}{135012} (\bibinfo{year}{2015}), \eprint{1410.3835}.

\bibitem[{\citenamefont{Jim\'enez-Forteza
  et~al.}(2017)\citenamefont{Jim\'enez-Forteza, Keitel, Husa, Hannam, Khan, and
  P\"urrer}}]{UIB2016_FF_paper}
\bibinfo{author}{\bibfnamefont{X.}~\bibnamefont{Jim\'enez-Forteza}},
  \bibinfo{author}{\bibfnamefont{D.}~\bibnamefont{Keitel}},
  \bibinfo{author}{\bibfnamefont{S.}~\bibnamefont{Husa}},
  \bibinfo{author}{\bibfnamefont{M.}~\bibnamefont{Hannam}},
  \bibinfo{author}{\bibfnamefont{S.}~\bibnamefont{Khan}}, \bibnamefont{and}
  \bibinfo{author}{\bibfnamefont{M.}~\bibnamefont{P\"urrer}},
  \bibinfo{journal}{Phys. Rev. D} \textbf{\bibinfo{volume}{95}},
  \bibinfo{pages}{064024} (\bibinfo{year}{2017}),
  \urlprefix\url{https://link.aps.org/doi/10.1103/PhysRevD.95.064024}.

\end{thebibliography}

\end{document}